\documentclass[preprint,aps,epsfig,floatfix]{revtex4}
\usepackage{graphicx}
\usepackage{dcolumn}
\usepackage{bm}
\usepackage{amsmath}
\begin{document}

\title{Model for Vortex Pinning in a Two-Dimensional Inhomogeneous d-Wave Superconductor}

\author{Daniel Valdez-Balderas} \email{balderas@mps.ohio-state.edu}

\author{David Stroud} \email{stroud@mps.ohio-state.edu}

\affiliation{Department of Physics, The Ohio State University, Columbus,
  Ohio 43210}

\date{\today}

\begin{abstract}
  We study a model for the pinning of vortices in a two-dimensional,
  inhomogeneous, Type-II superconductor in its mixed state.  The model
  is based on a Ginzburg-Landau (GL) free energy functional whose
  coefficients are determined by the mean field transition temperature
  $T_{c0}$ and the zero temperature penetration depth $\lambda(0)$. We
  find that if (i) $T_{c0}$ and $\lambda(0)$ are functions of
  position, and (ii) $\lambda^{2}(0)\propto T_{c0}^{y}$ with $y>0$,
  then vortices tend to be pinned by regions where $T_{c0}$, and
  therefore the magnitude of the superconducting order parameter
  $\Delta$, are large.  This behavior is in contrast to the usual
  picture of pinning in Type-II superconductors, where pinning occurs in
  the small-gap regions.
  We also compute the local density of states of a model BCS
  Hamiltonian with d-wave symmetry, in which the pairing field
  $\Delta$ is obtained from the Monte Carlo simulations of a GL free
  energy.  Several features observed in scanning tunneling
  spectroscopy measurements on YBa$_2$Cu$_3$O$_{6+x}$ and
  Bi$_2$Sr$_2$CaCu$_2$O$_{8+x}$ are well reproduced by our model: 
  far from vortex cores, the local density of states (LDOS) spectra
  has a small gap and sharp coherence peaks, while near the vortex
  cores it has a larger gap with low, broad peaks.  Additionally, also
  in agreement with experiment, the spectra near the core does not
  exhibit a zero energy peak which is, however, observed in other
  theoretical studies.
\end{abstract}

\maketitle
\section{Introduction}

It is generally believed that vortices in Type II superconductors tend to be pinned
in regions where the gap is small~\cite{blatter}.  This is true because a vortex, being a region
where the gradient of the superconducting order parameter is large, locally increases the gradient
part of the free energy.  Since this local increase is less in regions where the gap is smaller than average,
the vortex tends to migrate to such regions, according to this picture.  

In this paper, we describe a simple model for vortex pinning in an inhomogeneous two-dimensional (2D) superconductor, in which the vortices tend to be pinned in regions where the gap is {\it larger} than its spatial average.  The model is based on a Ginzburg-Landau (GL) free energy functional in which both the mean-field transition temperature $T_{c0}$ and the zero-temperature penetration depth $\lambda(0)$ are functions
of position, but are correlated in such a way that regions with large $T_{c0}$ also have large $\lambda(0)$.  
This assumption seems to apply to some of the high-T$_c$ cuprate superconductors: according to scanning
tunneling microscopy (STM) experiments on cuprates~\cite{cren, pan_oneal_badzey, howald1,
lang_davis, howald2, kato_sakata, fang, mashima}, regions that have a large gap (proportional to
$T_{c0}$ in this model) also have a single-particle local density of states (LDOS) with low, broad peaks, suggestive of a low superfluid density in these regions [proportional to $1/\lambda^2(0)$].  Our model is a generalization of an earlier approach intended to treat inhomogeneous superconductors in zero magnetic 
field\cite{valdez_stroud2}.

In order to further test this vortex pinning model, we also examine the quasiparticle LDOS near the
vortex cores within this model.  The proper theoretical description of this LDOS near the cores
is one of the unsolved issues in the field of high-T$_c$ 
superconductivity.  STM experiments on 
YBa$_2$Cu$_3$O$_{6+x}$ (YBCO) \cite{maggio_renner_fischer,
  renner_revaz_fischer} and Bi$_2$Sr$_2$CaCu$_2$O$_{8+x}$ (Bi2212)
\cite{levy_kugler_fischer, pan_hudson_gupta, matsuba_sakata_kosugi,
  hoogenboom_renner_revaz}, show that the local density of states
(LDOS) near vortex cores in those materials has the
following characteristics: a dip at zero energy, small peaks at
energies smaller than the superconducting gap, and low, broad peaks at
energies above the superconducting gap.  Those features contrast with
the spectra shown by conventional superconductors near vortex
cores, where the LDOS usually has a peak at zero energy
in clean superconductors (i.e., those with a mean-free path larger
than the coherence length) \cite{hess_robinson_waszczak,
  renner_kent_niedermann}, or is nearly energy-independent in dirty superconductors
\cite{renner_kent_niedermann, wilde_iavarone_welp}. 

The zero-energy peak in the LDOS near vortex cores of
clean conventional superconductors can be understood in terms of
electronic states with subgap energies bound to vortex cores
\cite{caroli_gennes_matricon, shore_huang_dorsey}. However, the
structure of the spectra near vortex cores of cuprates is still
lacking an explanation.  Several authors have suggested that this
structure is due to some type of competing order which emerges within
the vortex cores when superconductivity is suppressed by a magnetic
field \cite{franz_tesanovic1, arovas_berlinsky_kallin,
  franz_tesanovic2, han_lee, kishine_lee_wen}.  Some of those models,
and other descriptions of the spectra near vortex cores, have used the
Bogoliubov-de Gennes method to solve various microscopic Hamiltonians
\cite{soininen_kallin_berlinsky, wang_macdonnald, franz_tesanovic1,
  zhu_ting, melikyan_tesanovic, vafek_melikyan_franz_tesanovic}, such
as BCS-like models with d-wave symmetry.  One recent model for the
LDOS near the vortex cores, proposed by Melikyan
and Tesanovic\cite{melikyan_tesanovic}, used a Bogoliubov-de Gennes
approach to a tight-binding Hamiltonian.  These authors found that, if
a homogeneous pairing field is used in the microscopic Hamiltonian,
the LDOS exhibits a zero-energy peak on the
atomic sites that are closest to the vortex cores; this peak is,
however, absent from all other sites near the vortex cores.  On the
other hand, they found that introducing an enhanced pairing strength
for electrons on nearest-neighbor atomic sites near the vortex cores
leads to a suppression of the zero energy peak, thus obtaining a better
agreement with experiment.  They speculate that the enhancement could
be due to impurity atoms that pin the vortices, to a distortion of the
atomic lattice by the vortex itself, or to quantum fluctuations of the
superconducting order parameter.   The assumption of an enhanced pairing strength
is consistent with the model that we describe in this paper.  


In our model, having introduced this correlation between the GL parameters, we
anneal the system to find both the magnitude and the phase of the
superconducting order parameter that minimize the GL free energy at
low temperatures.  Since the magnetic vector potential enters the GL
free energy functional, this procedure naturally leads to vortex
formation.  The vortex cores can be identified in our simulations as
regions with a large phase gradient. We find that, in inhomogeneous
systems, vortices tend to be pinned in regions where the
superconducting gap is large.  For comparison, we also perform a
similar annealing procedure for homogeneous systems.  In this case,
contrary to inhomogeneous systems, the magnitude of the
superconducting order parameter is reduced near vortex cores.
Thus, the assumption that the gap is large in regions with small
superfluid density in inhomogeneous systems, originally intended to
model superconductors in a zero magnetic field \cite{valdez_stroud2},
leads naturally to pinning of the vortices in large-gap regions.  This
approach might therefore be complementary to that of Ref.\ 
\cite{melikyan_tesanovic} mentioned above.

To connect our vortex pinning model to previous studies of the LDOS
near vortex cores, we have also studied a microscopic 
Hamiltonian for electrons on a lattice.  This is a tight-binding model
in which electrons on nearest neighbor sites experience a pairing
interaction of the BCS type with d-wave symmetry \cite{valdez_stroud2,
  eckl1}. The LDOS is obtained by exact numerical diagonalization of
this Hamiltonian.  We take the pairing stregth between electrons on 
nearest neighbor sites to be
proportional to the value of the superconducting order parameter as
determined by the GL simulations using the GL functional just
described.

Using this combination of a GL free energy functional and a
microscopic d-wave BCS Hamiltonian, we find that a number of
experimental results on cuprates are reproduced well by our model for
inhomogeneous systems.  For example, far from vortex cores, the LDOS
shows sharp coherence peaks.  Near the vortex cores, our calculated
LDOS does not show a spurious zero energy peak; instead it exhibits a
large gap, as well as low, broad peaks which occur at energies larger
than the value of the superconducting energy gap observed far from
vortex cores.  Also, in agreement with experiment,
the LDOS curves near vortex cores are similar
to those in the large gap regions in systems
with quenched disorder but {\em zero} magnetic field\cite{valdez_stroud2}); 
this connection is discussed in
section \ref{results}.  One feature not captured by our model is the
existence of small, low energy peaks in the LDOS observed near the
vortex cores.

The rest of the present article is organized as follows.  In Section
\ref{model}, we present the GL free energy functional, as well the
microscopic Hamiltonian.  In Section \ref{results}, we present results
for homogeneous and inhomogeneous systems in a magnetic field, as well
as results for inhomogeneous systems in a zero magnetic field for
comparison.  Finally, in Section \ref{discussion}, we conclude with a
discussion and a summary of our work.

\section{Model}
\label{model}

\subsection{Ginzburg-Landau free energy functional}

We use a model for a single layer of a cuprate superconductor in a
perpendicular magnetic field based on a GL free energy functional of
the form described previously\cite{valdez_stroud2}:
\begin{eqnarray}
    \frac{F}{K_{1}} =
    \sum_{i=1}^{M} \left(\frac{t}{t_{c0i}}+3\right)\frac{1}{\lambda^{2}_{i}(0)t_{c0i}^2} \left|\psi_{i}\right|^2 +
    \sum_{i=1}^{M} \frac{1}{2(9.38)}\frac{1}{\lambda^{2}_{i}(0)t_{c0i}^4} \left|\psi_{i}\right|^4  \nonumber \\
    -\sum_{\langle ij \rangle}
    \frac{2|\psi_{i}||\psi_{j}|}{\lambda_{i}(0)t_{c0i}\lambda_{j}(0)t_{c0j}}
    \cos(\theta_{i}-\theta_{j}+A_{\vec i \vec j}).
  \label{eq:GLgeneral4}
\end{eqnarray}
In Eq.~(\ref{eq:GLgeneral4}) the first and second sums are
carried over $M$ square cells, each of area equal to the
zero-temperature GL coherence length squared, $\xi_{0}^2$, of a square
lattice into which the superconductor has been discretized for
computational purposes. The third sum is carried out over nearest
neighbor cells $\langle ij \rangle$.  Here
\begin{equation}
  \psi_{i} \equiv \frac{\Delta_{i}}{E_{0}},
  \label{eq:psi_def}
\end{equation}
where $\Delta_{i}$ is the complex superconducting order parameter of
the i$^{th}$ cell, $E_{0}$ is an arbitrary energy scale 
(which we take to be the hopping constant $t_{hop}\sim 200$ meV)
\begin{equation}
  t \equiv \frac{k_{B}T}{E_{0}},
  \label{eq:t_def}
\end{equation}
is the reduced temperature, $T$ is the temperature, and $k_{B}$ is the
Boltzmann constant.  Also, $\lambda_{i}(0)$ is the $T = 0$ penetration
depth and $t_{c0i}\equiv k_{B}T_{c0i}/E_{0}$, where $T_{c0i}$ is the
mean-field transition temperature of the i$^{th}$ cell. Therefore, in discretizing the
superconductor, we have assumed that $\lambda(0)$, $T_{c0}$ and
$\Delta$ are constant over distances of order $\xi_{0}$.  Finally,
\begin{equation}
  A_{\vec i \vec j} = \frac{2 e}{\hbar c}\int_{\vec i}^{\vec j}\vec A(\vec r)\cdot d\vec r
  \label{eq:vec_pot1}
\end{equation}
is the integral of the vector potential $\vec A(\vec r)$ from the
center of cell $i$, located at $\vec i$, to the center of cell $j$,
located at $\vec j$, and  
$ 
K_{1} \equiv \hbar^4 d /[32 (9.38) \pi m^{\ast 2} \mu_{B}^{2}]
$,
where $\mu_{B}^2\simeq 5.4\times 10^{-5}$ eV-\AA$^3$ is the
square of the Bohr magneton, $m^{\ast}$ is twice the mass of a free
electron, $e$ is the absolute value of its charge, and $d$ is the
thickness of the superconducting layer.
In all of our simulations we use periodic boundary conditions. A gauge
for the magnetic field that allows this is given in
Ref.~\cite{valdez_stroud2, wenbin_lee_stroud} [see, e.\ g., eq.\ (51) of Ref.\
\cite{valdez_stroud2}, which gives this gauge choice explicitly for $N_v = 1$].


We note that the connection between the local superfluid density $n_{s,i}(T)$ and 
penetration depth $\lambda_i(T)$ is such that $n_{s,i}(0) \propto 1/\lambda_i^2(0)$~\cite{valdez_stroud2}.
Thus, the quantity $1/\lambda^2$ in eq.\ (1) is a way of describing the local superfluid density,
which varies over a length scale of $\xi_0$. 


We will be using the above free energy functional at both $T = 0$ and finite $T$.  Although
we call this a ``Ginzburg-Landau free energy functional,'' this name is really a misnomer, since the original GL functional was intended to be applicable only near the mean-field transition temperature. Strictly speaking, the correct free energy functional near $T = 0$ should not have the GL form, but would be expected to contain additional terms, such as higher powers of $|\psi|^2$.  We use the GL form for convenience, and because we expect it will exhibit the qualitative behavior, such as vortex pinning in the large-gap regions, that would be seen in a more accurate functional.  

We assume that the magnetic field is uniform, which is a good
approximation for cuprate superconductors in their mixed state, provided
the external field is not too close to the lower critical field
$H_{c1}$.  In the cuprates, the approximation is satisfactory because the
penetration depth is of the order of thousands of \AA, while the
intervortex distance for the fields we consider is
$\sim 100$~\AA.  We employ a gauge that permits
periodic boundary conditions, such that the flux through the lattice
can take any integer multiple of $hc/e$ \cite{valdez_stroud2,
  wenbin_lee_stroud}.  Thus, the number $N_v$ of flux quanta $hc/(2e)$ 
must be an integer multiple of two.

The procedure for choosing the parameters $\lambda_{i}(0)$ and
$t_{c0i}$ is similar to that used in Ref.~\cite{valdez_stroud2}.
Basically, in most of the system ($\alpha$ regions), 
we take $\lambda_{i}(0) \sim \lambda(0)$, where $\lambda(0)$ is the
in-plane penetration depth of a bulk cuprate superconductor
[$\lambda(0)\sim 1800$~\AA\ in Bi2212, for example], while $t_{c0i}$ is
determined from the typical energy gap in the LDOS as
observed in STM experiments. However, we will
also introduce regions ($\beta$ regions) in which $t_{c0}$ and
$\lambda(0)$ are larger than those bulk values.  Throughout this
article, when we refer to a ``gap in the LDOS,'' we mean the distance
between the two peaks in the LDOS spectra.  This is the same
gap definition used in Ref.\ \cite{lang_davis}.

In Ref.~\cite{valdez_stroud2}, we generally introduced 
inhomogeneities in the superconducting order parameter $\Delta_i$ by
assuming a binary distribution of $t_{ci0}$, randomly distributed in space: 
$\alpha$ cells with a small $t_{ci0}$ and
$\beta$ cells with a large $t_{ci0}$.  We also assumed a
correlation between $\lambda_{i}(0)$ and $t_{c0i}$ of the form
\begin{equation}
  \lambda_i^2(0) =  \lambda^{2}(0) \left(\frac{t_{c0i}}{t_{c0}}\right)^{y},
  \label{eq:lambda2_correl}
\end{equation}
with $y=1$. More generally, Eq.~(\ref{eq:lambda2_correl}),
with $y>0$, accounts for the fact that in STM experiments, regions
with a large gap seem to have a small superfluid density (low and
broad peaks.)

In the present article, instead of randomly distributed $\beta$
cells, we introduce two square regions with only $\beta$ cells, while
the rest of the lattice is assumed to have only $\alpha$ cells.  We
make this choice to study the effects of these
inhomogeneities on field-induced vortices.  We also consider a more
general model than Ref.\ \cite{valdez_stroud2}, allowing $y$ to
have values other than only $y = 1$.

As shown in Eq.\ (36) of Ref.~\cite{valdez_stroud2}, in the absence of
thermal fluctuations the coupling constant $J_{XY,ij}$ between cells
$i$ and $j$ is approximately
\begin{equation}
  J_{XY,ij}(t) \simeq \frac{2(9.38)\sqrt{ (1-t/t_{c0i})  (1-t/t_{c0j}) } }{\lambda_{i}(0)\,\lambda_{j}(0)}K_{1}.
  \label{eq:JXY1}
\end{equation}
(The factor of $K_1$ is missing in Ref.\ \cite{valdez_stroud2}.)
If $t<<t_{c0i}$ and $t <<
t_{c0j}$, then
\begin{equation}
  J_{XY,ij}(t) \simeq \frac{2(9.38)K_{1}}{\lambda_{i}(0)\,\lambda_{j}(0)}.
  \label{eq:JXY1_lowt}
\end{equation}
If we further choose a binary distribution of $t_{c0i}$, that is,
\begin{equation}
  t_{c0i} =
  \begin{cases}
    t_{c0}, & \text{if $i$ is on an $\alpha$ cell},\\
    ft_{c0}, & \text{if $i$ is on a $\beta$ cell},
  \end{cases}
  \label{eq:tc0_alpha_beta}
\end{equation}
where $f$ is any positive number (typically $f > 1$), then 
\begin{equation}
  J_{XY,ij}(t) \simeq \frac{2(9.38)K_{1}}{\lambda^2(0)}\cdot
  \begin{cases}
    1 & \text{if $i$ and $j\in \alpha$,}\\
    \frac{1}{f^{\frac{y}{2}}} & \text{if $i\in \alpha$ and $j\in \beta$ or if $i\in \beta$ and $j\in \alpha$,}\\
    \frac{1}{f^{y}} & \text{if $i$ and $j\in \beta$.}
  \end{cases}
  \label{eq:J_alpha_beta}
\end{equation}
This expression shows that in regions with a large gap the coupling
between XY cells is small, reflecting the large penetration depth in
those regions.

\subsection{Microscopic Hamiltonian}

Besides using a GL free energy to explore vortex pinning in an
inhomogeneous superconductor, we have also studied the LDOS of
a corresponding microscopic model Hamiltonian, given by~\cite{valdez_stroud2}:
\begin{equation}
  H =   2 \sum_{\langle i,j \rangle,\sigma}t_{ij}c_{i\sigma}^{\dagger}c_{j\sigma}
+2\sum_{\langle i,j \rangle}(\Delta_{ij}c_{i\downarrow}c_{j\uparrow} +
\text{c.c.}) -\mu \sum_{i,\sigma}c_{i\sigma}^{\dagger}c_{i\sigma}
  \label{eq:hamil_bcs}
\end{equation}
Here $\sum_{\langle i, j \rangle }$ denotes a sum over distinct pairs
of nearest neighbor atomic sites on a square lattice with $N$ sites,
$c_{j\sigma}^{\dagger}$ creates an electron with spin $\sigma$
($\uparrow$ or $\downarrow$) at site $j$, $\mu$ is the chemical
potential, and $\Delta_{ij}$ denotes the strength of the pairing
interaction between electrons at sites $i$ and $j$.  Finally, we write
$t_{ij}$ as 
\begin{equation}
  t_{ij} = -t_{hop}\, e^{-i A_{\vec i \vec j}^{\prime}},
  \label{hopp_const2}
\end{equation}
with
\begin{equation}
  A_{\vec i \vec j}^{\prime} = \frac{e}{\hbar c}\int_{\vec i}^{\vec j}\vec A(\vec r)\cdot d\vec r.
\end{equation}
Here the integral runs along the line from atomic site $i$, located at $\vec
i$, to the atomic site $j$, located at $\vec j$, and $t_{hop}>0$ is
the hopping integral for nearest neighbor sites on the lattice. Note
that the prefactor in $A_{ij}^{\prime}$ involves the factor of $hc/e$
due to a single electronic charge, and is thus twice as large as
that in $A_{ij}$, which involves the charge of a Cooper pair.

Following Ref.~\cite{valdez_stroud2} we take
$\Delta_{ij}$ to be given by
\begin{equation}
  \Delta_{ij} = \frac{1}{4}\frac{|\Delta_i|+|\Delta_j|}{2} e^{i
  \theta_{ij}},
  \label{delta}
\end{equation}
where
\begin{equation}
  \theta_{ij} =
  \begin{cases}
    (\theta_i+\theta_j)/2, & \text{if bond $\langle i, j \rangle$ is in $x$-direction,}\\
    (\theta_i+\theta_j)/2 + \pi, & \text{if bond $\langle i, j \rangle$ is in $y$-direction,}
  \end{cases}
  \label{eq:thetaij}
\end{equation}
and
\begin{equation}
  \Delta_{j}=|\Delta_j|e^{i\theta_{j}},
  \label{eq:psi_complex}
\end{equation}
is the value of the complex superconducting order parameter at site
$j$.  We will refer to the lattice over which the sums
in~(\ref{eq:hamil_bcs}) are carried out as the {\em atomic} lattice
(in order to distinguish it from the {\em XY} lattice.) The first term
in Eq.~(\ref{eq:hamil_bcs}) thus corresponds to the kinetic energy,
the second term is a BCS type of pairing interaction with $d$-wave
symmetry, and the third is the energy associated with the
chemical potential.

The model we present is similar to the one presented in
Ref.~\cite{valdez_stroud2}, the main differences being the inclusion of
a vector potential in the GL free energy functional, and the spatial
distribution of the inhomogeneities. 
Because the vector potential 
introduces vortices in the system, our results 
differ substantially from our previous work.

\section{Results}
\label{results}

\begin{figure}[htbp]\centering
  {\includegraphics[width=12cm]{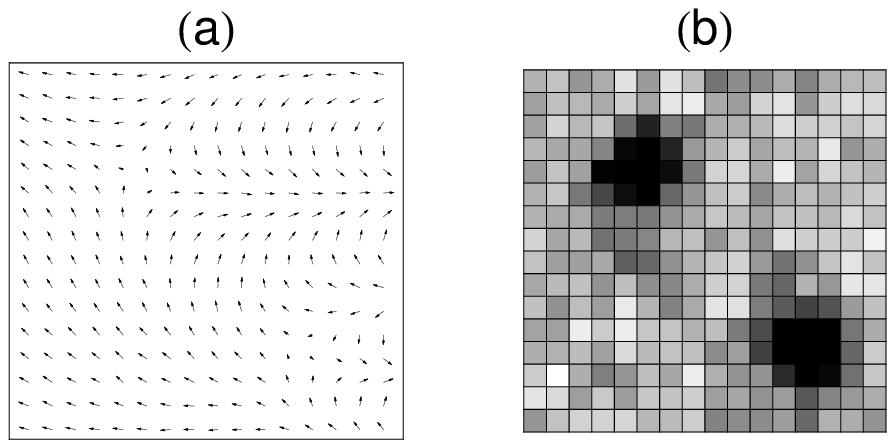}}
  \caption{
    Map of the pairing field $\Delta$ over a homogeneous,
    two-dimensional superconductor of area $16\xi_{0}\times 16
    \xi_{0}$ in a transverse, uniform magnetic field $B\simeq 6$T.
    The superconductor has been discretized into XY cells, each of
    which has an area $\xi_{0}^2$ and encloses nine atomic sites. In
    part (a) the length and direction of each arrow represents the complex value of $\Delta$
    within an XY cell.  Part (b) shows a map of the magnitude $|\Delta|$ of the
    pairing field.  Dark (light) regions represent a small (large)
    value of $|\Delta|$.  Vortex cores can be easily identified as the
    darkest regions.}
  \label{fig:psi2_L16_T001_ktc14_f1}
\end{figure}

We first present results for homogeneous systems in the presence of an
applied transverse magnetic field equal to two flux quanta through the lattice at low $T$. 
This is the lowest magnetic field consistent with the periodic boundary conditions.
In this case,
$t_{c0i}$ and $\lambda_{i}(0)$ are independent of $i$.
Fig.~\ref{fig:psi2_L16_T001_ktc14_f1} shows our calculated maps of $\Delta$ for
a homogeneous system.  In part
(a), the lengths and directions of the arrows represent
the magnitude and phase of $\Delta$ in
each XY cell.  Even though $t_{c0i}$ is homogeneous,
the magnetic field renders $\Delta$ inhomogeneous, especially near the
vortex cores, where $\Delta$ has a large phase gradient and
a smaller magnitude.  This behavior is familiar from Ginzburg-Landau
treatments of homogeneous Type II superconductors in a magnetic field.
Part (b) of Fig.~\ref{fig:psi2_L16_T001_ktc14_f1} shows a map of
$|\Delta|$, with dark (light) regions representing small (large) value
of $|\Delta|$.  The vortex cores are the
darkest regions.

\begin{figure}[htbp]\centering
  {\includegraphics[width=12cm]{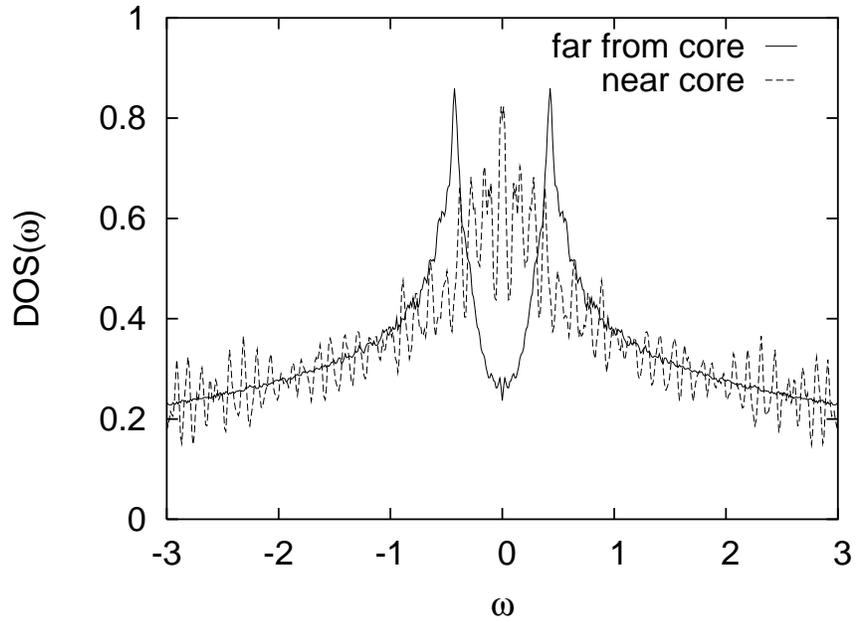}}
  \caption{
    Local density of states averaged over two different groups of XY
    cells in the homogeneous superconductor described in
    Fig.~\ref{fig:psi2_L16_T001_ktc14_f1}: far from the vortex core
    (full curve) and near the core (dotted curve).  The oscillations
    in the dotted curve are due
    to numerical fluctuations arising from the small number of sites
    in the core regions}
  \label{fig:ldos_L16_na1_f1}
\end{figure}

Fig.~\ref{fig:ldos_L16_na1_f1} show the LDOS averaged over regions
near and far from the vortex cores of the system described in
Fig.~\ref{fig:psi2_L16_T001_ktc14_f1}. Since $|\Delta|$ is small near
the core, we have defined an XY cell to be ``near the core'' if
$|\Delta| \leq |\Delta_{\text{avg}}|/2$ in that cell, where
$|\Delta_{\text{avg}}|$ is the value of $|\Delta|$ averaged over all
the XY cells of the lattice. All other cells are considered to be far
from the core. Thus, in order to compute the averaged quantities shown
in Fig.~\ref{fig:ldos_L16_na1_f1}, we first calculated the LDOS on every
atomic site by exact numerical diagonalization of the Hamiltonian
(\ref{eq:hamil_bcs}), and then averaged the LDOS over the the set of
atomic sites near and far from the vortex cores, as defined
above.  We have chosen to enclose nine atomic sites inside
each XY cell for this system, because the
coherence length in cuprates is $\sim 15$~\AA, approximately
three times larger than the atomic lattice constant, $\sim 5$~\AA.

Fig.~\ref{fig:ldos_L16_na1_f1} shows that, in a homogeneous system,
the LDOS far from the core is strongly suppressed near $\omega=0$,
and exhibits sharp coherence peaks, reminiscent of a d-wave
superconductor in the absence of a magnetic field.  Near the vortex
cores, on the other hand, the gap is filled, and the LDOS is large
near $\omega=0$, resembling the spectrum of a gapless tight-binding
model in two dimensions and zero magnetic field \cite{assaad,valdez_stroud2}.
The spectrum near the core has considerable numerical noise, because
it represents an average over only a few atomic sites.  In particular, the
rapid oscillations are probably due to this numerical artifact.

To estimate our magnetic fields, we note that,
because of the numerical implementation of the periodic boundary
conditions \cite{yu_stroud,valdez_stroud2}, the field has to be chosen
so that the number $N_v$ of magnetic flux quanta $\Phi_{0} = hc/(2e)$
through the system is a multiple of two.  Therefore, the magnetic
field can be estimated using $B=N_v \Phi_{0}/S$ where $S$ is the area
of the system, and $\Phi_{0}\equiv hc/(2e)\approx 2\times 10^{-15}$~T-m$^{2}$. 
In the present article we use a numerical sample of area
area $S=(48 a_{0})^{2}$, where $a_{0} \sim 5$ \AA \ is the atomic
lattice constant.  Therefore $B\approx N_{v}\times 3$~T, and for the
systems containing two vortices,
$B \sim 6$T.
The magnitude of this magnetic field is comparable
used in STM experiments\cite{levy_kugler_fischer,pan_hudson_gupta}.

\begin{figure}[htbp]\centering
  {\includegraphics[width=12cm]{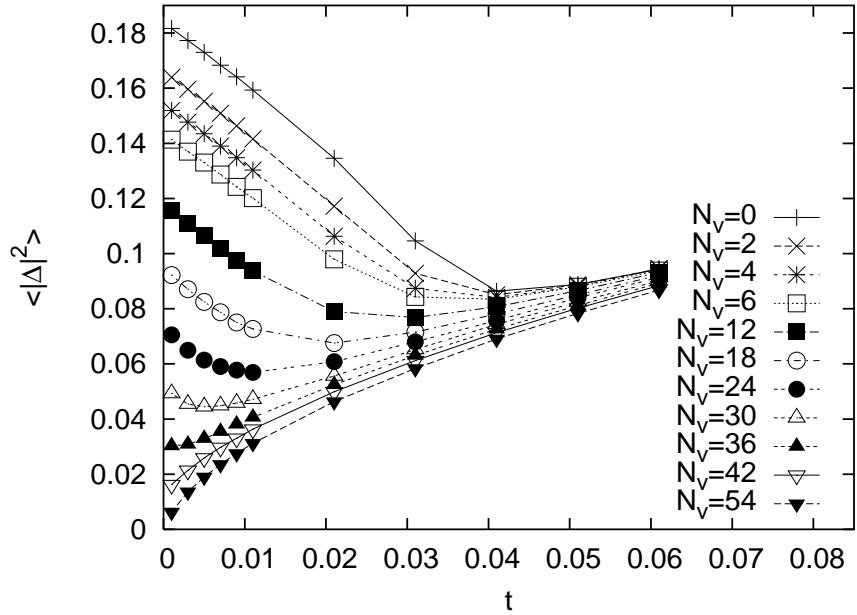}} 
  \caption{
    Thermal average $\langle |\Delta|^2 \rangle$ of the squared
    magnitude of the superconducting order parameter versus the
    reduced temperature $t$ for a system with a homogeneous $t_{c0}$
    placed in various fields. The magnitude of the magnetic field is
    $B\simeq N_{v}\times 3$T as described in the text.  
    We can observe that the minimum of $\langle |\Delta|^2\rangle(t)$
    versus $t$ (which is known to occur near the phase ordering
    temperature in zero magnetic field systems
    \cite{bittner_janke,valdez_stroud2}) is shifted toward smaller
    values of $t$ with increasing magnetic field.}
  \label{fig:psi2_gl_L16} 
\end{figure}

We now briefly discuss the temperature evolution of $\langle
|\Delta|^2 \rangle$, defined as an average of $\langle |\Delta_i|^2
\rangle$ over all XY lattice cells $i$, for systems with homogeneous
$t_{c0}$ in a transverse magnetic field. Here, $\langle... \rangle$
denotes a thermal average, obtained using Monte Carlo simulations.
Fig.~\ref{fig:psi2_gl_L16} shows curves of $\langle |\Delta|^2
\rangle(t)$ versus reduced temperature $t$, for systems of the same
size subject to different magnetic fields, and therefore containing
different numbers of vortices $N_v$.  In the zero magnetic field case,
$N_v = 0$, $\langle |\Delta|^2\rangle(t)$ has a minimum as a function
of $t$, which occurs near the phase ordering temperature.
Fig.~\ref{fig:psi2_gl_L16} shows that, as the field is increased, the
phase ordering temperature is reduced, and around $N_v = 36$, it seems
to drop to zero. The fact that of $\langle |\Delta|^2 \rangle(t) $
increases with $t$ for large $t$ is an artifact of the GL functional,
as has been discussed in detail in \cite{valdez_stroud2} for the case
$B = 0$.

\begin{figure}[htbp]\centering

  {\includegraphics[width=16cm]{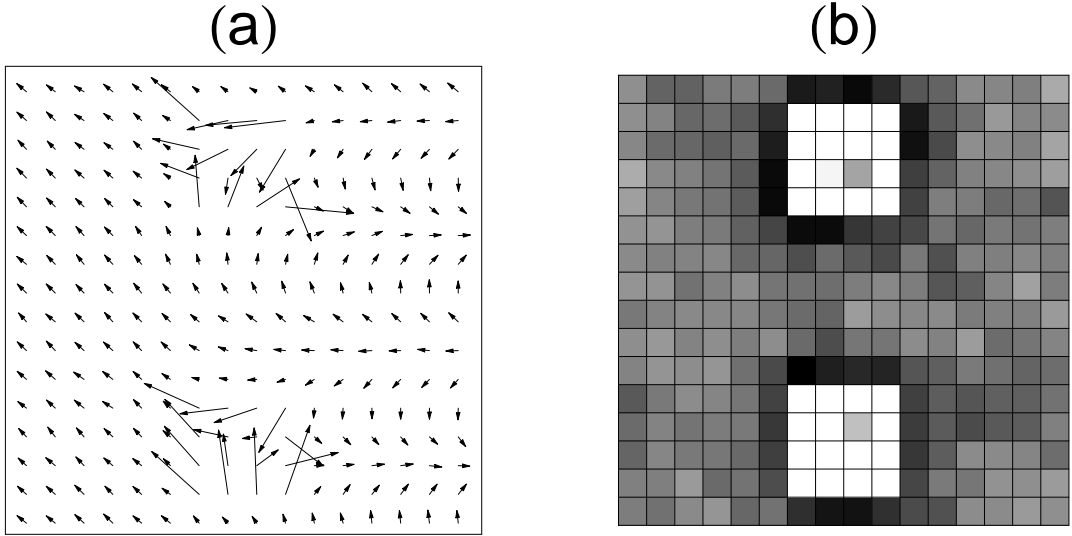}}
  \caption{
    Map of the pairing field $\Delta$ in  an inhomogeneous,
    two-dimensional superconductor of area $16\xi_{0}\times 16
    \xi_{0}$ in a transverse, uniform magnetic field $B\simeq 6$ T.
    The superconductor has two $\beta$ regions where $t_{c0i}$ is
    large; these correspond to the light regions in (b).  The system
    has been discretized into XY cells, each of which has an area
    $\xi_{0}^2$ and encloses nine atomic sites. In part (a) each arrow
    represents the complex value of $\Delta$ within an XY cell.  Part
    (b) shows a map of the magnitude $|\Delta|$ of the pairing field.
    Dark (light) regions represent a small (large) value of
    $|\Delta|$.  The locations of the two vortex cores can be
    identified as the regions with a large phase gradient in part (a).
    They are pinned to regions with a large $|\Delta|$ because of the
    low value of the coupling between XY cells in those regions. The
    present results are obtained from Eqs.\ (\ref{eq:tc0_alpha_beta}) and
    (\ref{eq:J_alpha_beta}), using $f=3$ and $x=3$.}
  \label{fig:psi2_L16_T001_ktc14_f1_fac3_na3_x3}
\end{figure}

We now proceed to show systems with inhomogeneities.
Fig.~\ref{fig:psi2_L16_T001_ktc14_f1_fac3_na3_x3}~shows gap maps of
systems in which $t_{c0i}$, and therefore $J_{XY,ij}$, are are $i$
dependent. Specifically, two square regions, of size 4$\times$4 XY
cells each, have $t_{c0i} = ft_{c0}$, with $f=3$; those are called
$\beta$ cells, as described in the previous section. The remainder of
the XY cells have $t_{c0i}=t_{c0}$, and are called $\alpha$ cells.  
Clearly, the vortices are pinned in the $\beta$ regions, where
$|\Delta|$ is large. The fact that $|\Delta|$ is large in $\beta$
regions is of course due to the fact that at low temperatures,
$|\Delta|$ is roughly proportional to $t_{c0}$.
On the other hand, in order to understand why the vortices are pinned
in the large-$|\Delta|$ regions, we note that $J_{XY,ij}$ at low
temperatures can be estimated with the use of eq.\ 
(\ref{eq:J_alpha_beta}). For the particular parameters used in this
calculation, namely, $x=3$ and $f=3$, $J_{XY,ij}$ is about 27 times
smaller within $\beta$ regions than in the $\alpha$ regions.  Because
of this ratio, phase gradients near the vortex cores cost much less energy
in the $\beta$ regions than in the $\alpha$ regions, even though
$|\Delta|$ is larger in the $\beta$ regions.  Thus, it is
energetically favorable for the vortices to be pinned in the $\beta$
regions.

\begin{figure}[htbp]\centering
  {\includegraphics[width=11cm]{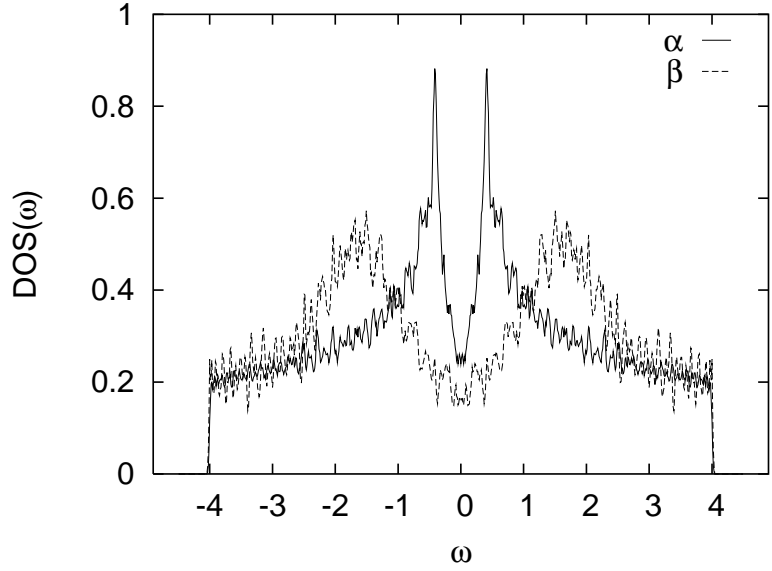}}
  \caption{
    Local density of states averaged over two different groups of XY
    cells in the inhomogeneous superconductor shown in
    Fig.~\ref{fig:psi2_L16_T001_ktc14_f1_fac3_na3_x3}: far from the
    core ($\alpha$) and near the core ($\beta$). In agreement with
    experimental results, this Figure shows that (i) far from vortex
    cores the LDOS shows sharp coherence peaks; (ii) near the vortex
    cores, the LDOS does not display the unphysical zero-energy peak
    obtained in other models.  Instead (iii), it has a large gap, as
    well as low and broad peaks.
  }
  \label{fig:ldos_L16_t0001_tc014_f1_na3_x3}
\end{figure}

Fig.~\ref{fig:ldos_L16_t0001_tc014_f1_na3_x3} shows the LDOS averaged
over $\alpha$ and $\beta$ regions of the system shown in
Fig.~\ref{fig:psi2_L16_T001_ktc14_f1_fac3_na3_x3}.  This Figure shows
that, far from the vortex cores ($\alpha$ regions), the LDOS is very
similar to that of regions far from the core in homogeneous system at
low magnetic field [cf.\ Fig.\ \ref{fig:ldos_L16_na1_f1}] - both
spectra have a small gap and sharp coherence peaks.  However, in
contrast to the homogeneous case, the LDOS of the inhomogeneous system
exhibits a large gap and broadened peaks near the vortex core.  This
LDOS spectrum is similar to that calculated for the large gap regions of
systems with quenched disorder and no magnetic field in our previous
work \cite{valdez_stroud2}. In the present case, the gap in the LDOS
is large in the $\beta$ regions, of course, because of the large value
of $t_{c0i}$ and therefore of $|\Delta|$, near the vortex cores.  The
peaks in the LDOS in the $\beta$ region above the gap are low and
broad, on the other hand, because of the large phase gradient of the
superconducting order parameter in these regions.  It is as if the
system has lost phase coherence in those regions due to the presence
of the vortex core.  We can more clearly see this point by comparing
the LDOS of a $\beta$ region containing a pinned vortex core to that
of a $\beta$ region in a system in a zero magnetic field, in which,
therefore, there is no vortex core to be pinned. We now describe this
system.

\begin{figure}[htbp]\centering

  {\includegraphics[width=15cm]{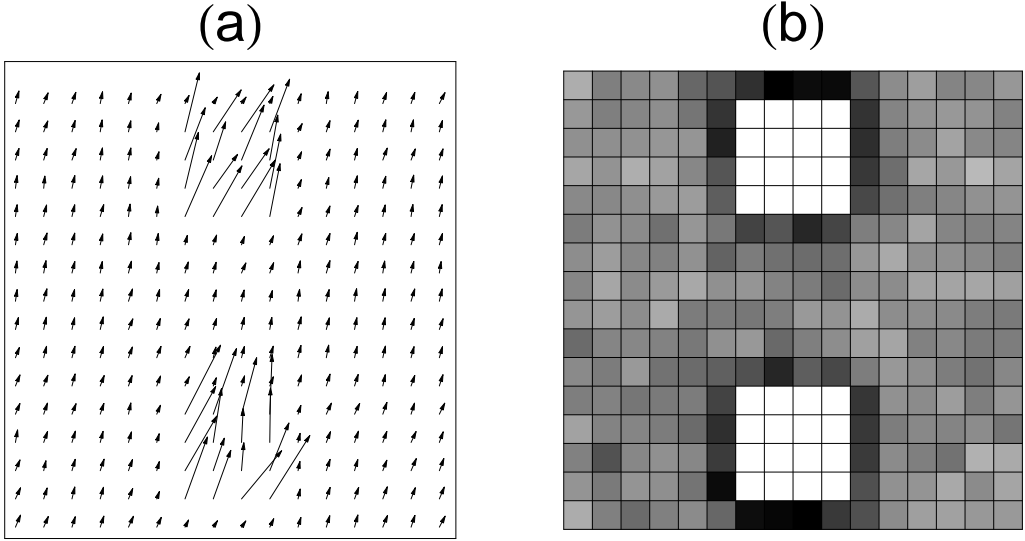}}
  \caption{
    Same as Fig.~\ref{fig:psi2_L16_T001_ktc14_f1_fac3_na3_x3}, but for 
    a system at $B = 0$ instead of $B \neq 0$.   At low
    temperatures, the phase of $\Delta$ is almost uniform, since, in
    the absence of a magnetic field, phase gradients cost energy.  As
    expected, $\Delta$ is larger in the $\beta$ regions, which have a
    large value of $t_{c0i}$. }
  \label{fig:psi2_L16_T001_ktc14_f1_fac3_na3_f0xy}
\end{figure}

\begin{figure}[htbp]\centering
  {\includegraphics[width=12cm]{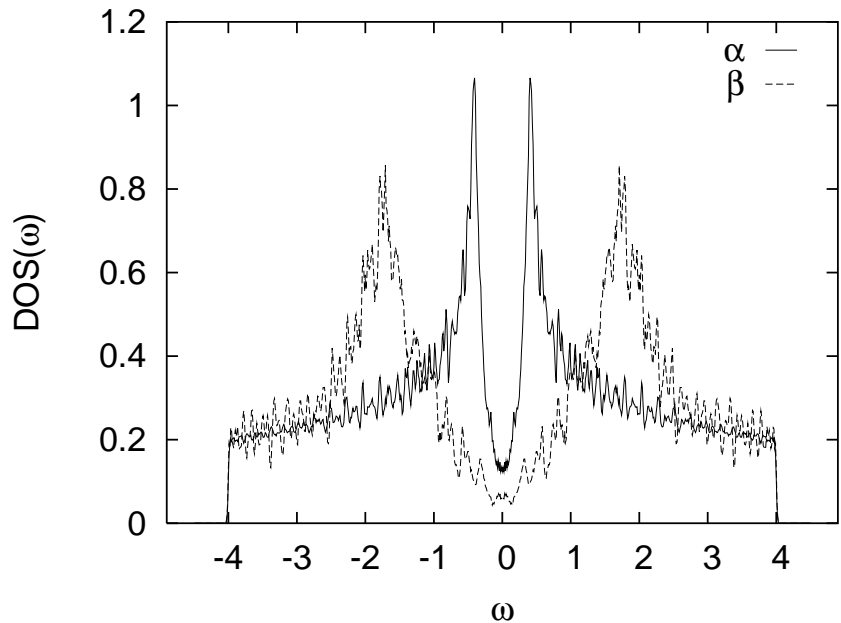}}
  \caption{
    Local density of states for the system shown in
    Fig.~\ref{fig:psi2_L16_T001_ktc14_f1_fac3_na3_f0xy}. The LDOS in
    the $\beta$ regions of this system has much sharper peaks than the
    LDOS of the $\beta$ regions for the corresponding system in a
    finite magnetic field shown in
    Fig.~\ref{fig:ldos_L16_t0001_tc014_f1_na3_x3}.  This result shows
    that having a large gap is not a sufficient condition to observe
    broadened peaks near vortex cores; a large phase gradient is also
    required.}
  \label{fig:ldos_L16_t0001_tc014_f0xy_na3_x3}
\end{figure}

Fig.~\ref{fig:psi2_L16_T001_ktc14_f1_fac3_na3_f0xy} shows a system at
low temperature with two inhomogeneities, similar to that shown in
Fig.~\ref{fig:psi2_L16_T001_ktc14_f1_fac3_na3_x3}, but with zero
rather than a finite magnetic field.  Clearly, in the ground state, as
expected, the phase of $\Delta$ is almost uniform. Furthermore, and
also as expected, $|\Delta|$ is larger in the $\beta$ regions, which
have a larger value of $t_{c0i}$.  However, when we turn to the LDOS
(Fig.~\ref{fig:ldos_L16_t0001_tc014_f0xy_na3_x3}), the LDOS of the
$\beta$ regions is characterized by much sharper coherence peaks than
that at finite magnetic field described in the previous paragraph,
presumably because of the absence of a large phase gradient.

\begin{figure}[htbp]\centering
  {\includegraphics[width=15cm]{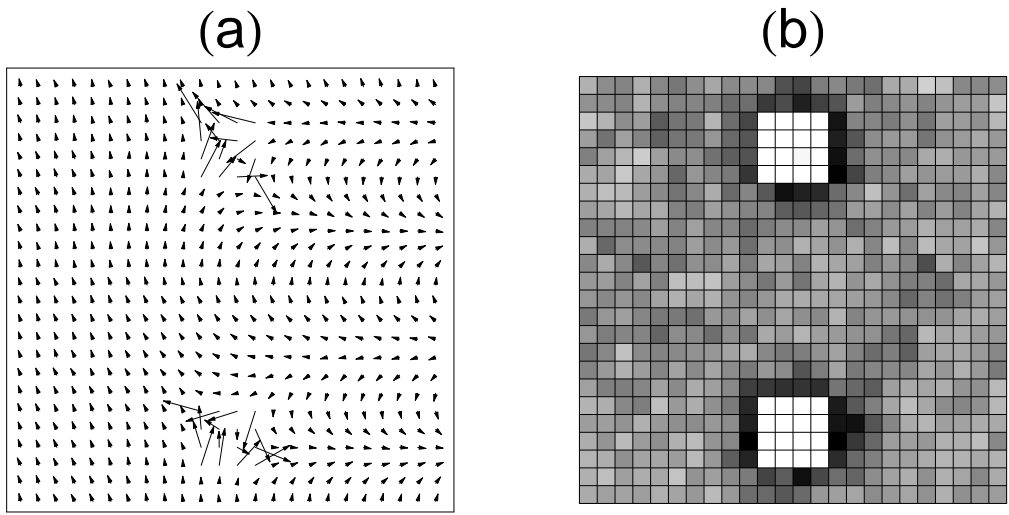}}
  \caption{
    Similar to Fig.~\ref{fig:psi2_L16_T001_ktc14_f1_fac3_na3_x3}, but
    the system has 24$\times$24 instead of 16$\times$16 XY cells, each with
    four instead of nine atomic lattice sites.}
  \label{fig:psi2_L24_T001_ktc14_f1_fac3_na2_x3}
\end{figure}

\begin{figure}[htbp]\centering
  {\includegraphics[width=12cm]{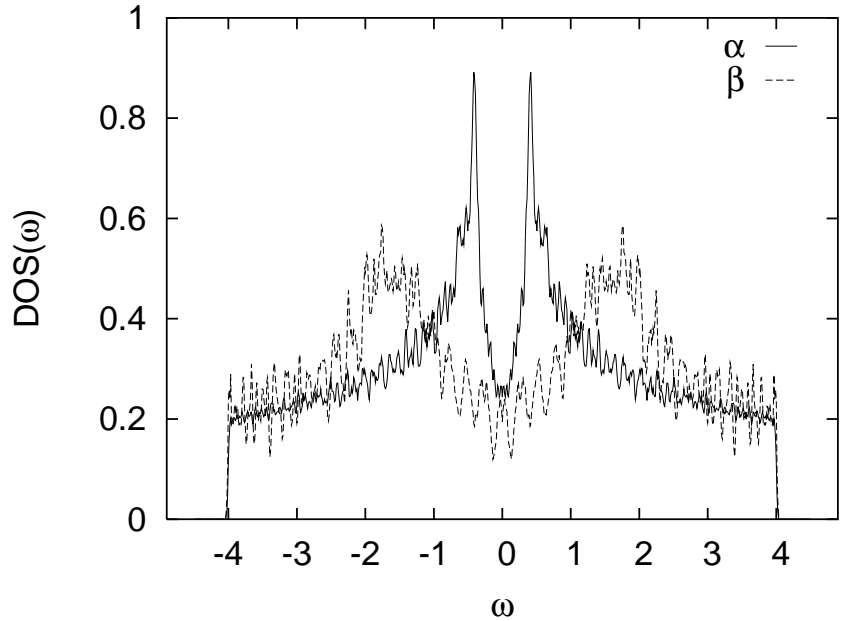}}
  \caption{
    Local density of states averaged over regions far from the vortex
    cores, for the system shown
    Fig.~\ref{fig:psi2_L24_T001_ktc14_f1_fac3_na2_x3}. Results are
    very similar to the corresponding system with 16$\times$16 cells, which
    has nine instead of four atoms per XY cell, and is shown in
    Fig.~\ref{fig:ldos_L16_t0001_tc014_f1_na3_x3}. Thus, our
    results are not strongly dependent on the size of the
    XY cell or of the XY lattice.}
  \label{fig:ldos_L24_t0001_tc014_f1_na2_x3}
\end{figure}

\begin{figure}[htbp]\centering
  {\includegraphics[width=15cm]{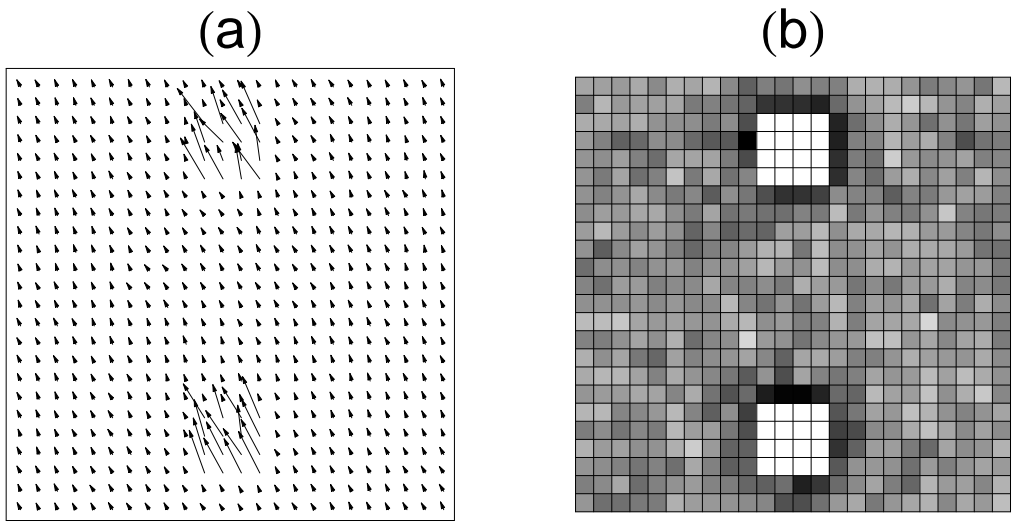}}
  \caption{
    Same as Fig.~\ref{fig:psi2_L16_T001_ktc14_f1_fac3_na3_f0xy} but
    for a system consisting of 24$\times$24 instead of 16$\times$16 XY
    cells, each having four instead of nine atomic sites.}
  \label{fig:psi2_L24_T001_ktc14_f1_fac3_na2_f0xy}
\end{figure}

\begin{figure}[htbp]\centering
  {\includegraphics[width=12cm]{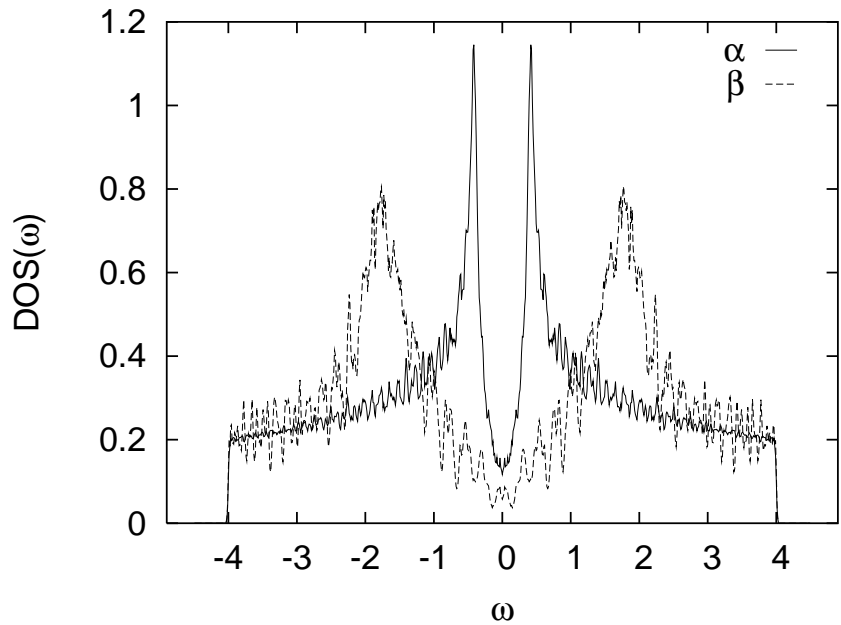}}
  \caption{
    Same as Fig.~\ref{fig:ldos_L16_t0001_tc014_f0xy_na3_x3} but for
    a system of 24$\times$24 instead of 16$\times$16 XY cells, each having
    cell has four rather than the nine atomic sites.
  }
  \label{fig:ldos_L24_t0001_tc014_f0xy_na2_x3}
\end{figure}

We have also tested the sensitivity of our results to lattice size and
to the number of atomic sites per XY cell.  To do this, we performed
calculations similar to the ones described above, but with
24$\times$24 instead of 16$\times$16 XY lattices, each XY cell
containing four instead of nine atomic lattice sites.
Fig.~\ref{fig:psi2_L24_T001_ktc14_f1_fac3_na2_x3} shows the results
for a 24$\times$24 XY lattice of an inhomogeneous system in the
presence of a magnetic field. As in the 16$\times$16 case, the
vortices are pinned in the regions with large $\Delta$ and small
$J_{XY,ij}$.  Fig.~\ref{fig:ldos_L24_t0001_tc014_f1_na2_x3} shows that
the LDOS for this system both near and far from the core is nearly the
same as that of the 16$\times$16 system shown in
Fig.~\ref{fig:ldos_L16_t0001_tc014_f1_na3_x3}.
Likewise, a system
with 24$\times$24 XY cells and two atomic sites per XY cell, but with
no magnetic field, is shown in
Fig.~\ref{fig:psi2_L24_T001_ktc14_f1_fac3_na2_f0xy}; the corresponding
LDOS averaged over regions near and far from the core is shown in
Fig.~\ref{fig:ldos_L24_t0001_tc014_f0xy_na2_x3}.  The LDOS shown in
this Figure has the same features as that of the 16$\times$16 system
shown in Fig.~\ref{fig:ldos_L16_t0001_tc014_f0xy_na3_x3}.
We thus conclude that our
results are not very sensitive to the lattice sizes used, nor to the
number of atomic sites per XY cell.

In the calculations described above, have assumed that $\lambda^2
\propto T_{c0}^y $.  If $y \geq 3$, then the vortices are consistently
pinned in the large gap region.  By contrast, if $y = 1$, we find the
vortices may or may not be pinned in the large gap region, depending
on where these regions are located within the computational lattice.
The vortices obviously go to the large gap region because of the
correlation between $\lambda^2$ and $T_{c0}$.
As mentioned earlier, this correlation originates in the fact large
$\lambda^2$ implies a small energy cost to introduce a gradient in the
phase of the order parameter.  Such a gradient must exist near a
vortex core; therefore, the vortex prefers to be in a region where
this gradient is energetically inexpensive.  Melikyan and Tesanovic
\cite{melikyan_tesanovic} discuss various other possible causes of
larger pinning in the large gap region; these include quantum
fluctuations, distortion of the atomic lattice by the vortices, and
the pinning of vortices by impurity atoms.  Our model could be viewed
as a special kind of such impurity pinning, in which the
``impurities'' are superconducting regions with a large gap and large
penetration depth.  We have also looked at how the size of the pinning
regions affects the pinning.  The pinning is generally more effective
for large pinning regions, probably because the vortex energy is
reduced by a larger amount if pinned in a region of large pinning
area, all other parameters being the same.

Finally, we note the similarity between our calculated LDOS
versus energy curves for regions near the vortex cores, and the
corresponding curves for large gap regions of inhomogeneous systems
with quenched disorder in a {\it zero} magnetic field, calculated in
our previous work \cite{valdez_stroud2}.  Those similarities have been
implied in several experimental papers.  For example, Lang {\it et
  al.}\cite{lang_davis} discussed the similarity between the spectra
of large-gap regions of inhomogeneous systems at low temperatures 
%
in a zero field
and those observed in the pseudogap regime of some cuprate
superconductors.  In the pseudogap region, the phase configuration is
disordered and therefore, possibly like the $\beta$ region at zero
field, both might be considered as ``normal'' regions.  Likewise,
Fischer {\it et al.}~\cite{fischer_kugler_maggio} have also noted the
similarity of the low-$T$ spectra near the vortex cores to spectra in
the pseudogap regime.  Once again, the similarity may arise because
both the interior of the vortex cores and the pseudogap region may be
considered as ``normal''.
%
%
Thus, both of these reports implicitly suggest a similarity between
the low-$T$ spectra in the large gap regions of a disordered system at
$B = 0$ and the corresponding spectra near vortex cores at finite $B$.

\section{Discussion}
\label{discussion}

We have presented a model for the pinning of vortices by
inhomogeneities in a two-dimensional type II superconductor at low
temperatures. The model is based on a GL free energy functional, and
it is inspired by our previous study \cite{valdez_stroud2} of the LDOS
in an inhomogeneous superconductor in zero magnetic field.
In our model, we have
proposed a GL free energy functional in which regions with a large
value of the superconducting order parameter have a large penetration
depth, as suggested by zero-field STM experiments on cuprates. Using an annealing process and Monte Carlo simulations, we have found that the pinning of vortices by those large-gap regions
emerges naturally from the functional form of our GL free energy.  
The vortices are attracted to the large-gap regions, in our model,
because the large penetration depth leads to a small coupling between cells in
those regions. Therefore, the phase of the
superconducting order parameter can more easily bend in these regions, and this in
turn allows the large-gap regions to accomodate a vortex more easily
than regions with a small gap.  By contrast, in the absence of
quenched inhomogeneities, minimization of the GL free energy
functional yields a spatial configuration in which the superconducting
order parameter is suppressed near vortex cores.


It is worth commenting further on the qualitative physics underlying the pinning of the vortices in the region of enhanced local $T_c$.  Basically, this pinning behavior occurs because, in our model, a locally enhanced Tc corresponds to a locally suppressed superfluid density.   This superfluid density is proportional to the local $1/\lambda^2$, and is related to the Ginzburg-Landau coefficients, as we have described earlier.  The vortices can be more easily accommodated in these large-$T_c$ regions because the large phase gradients which characterize the vortices cost less energy in such regions.  Although we describe the pinning in terms of the variation of $1/\lambda^2$, the pinning is not due to any kind of ``magnetic'' forces - the quantity $1/\lambda^2$ is just a way of describing the local superfluid density.   Thus, just as in conventional pinning, the vortices are attracted to regions of lower free energy.  The major difference is that the superfluid density (and $1/\lambda^2$) is proportional to $T_c$ in the present model, rather than being independent of, or inversely correlated with $T_c$ as in more conventional pinning.  This leads to pinning in a large-$T_c$ region rather than a small-$T_c$ region.  We also emphasize that, although our pinning results are obtained using rather elaborate numerical calculations, the underlying physics is straightforward: the pinning behavior is consistent with qualitative expectations, given the model.   

It may seem strange to consider the spatial variation in $1/\lambda^2$ as occurring over a scale
of $\xi_0 \sim 15 \AA$, when $\lambda$ itself is of order $10^3$ $\AA$ or more.  However, it should be remembered that $1/\lambda^2$ is related to the coefficients of the
Ginzburg-Landau free energy [see eq.\ (1)], and these coefficients are, in fact, expected to
vary over a scale of $\xi_0$.  Thus, the model is, indeed, reasonable and consistent with the expected physics.   It is the local superfluid density (proportional to $1/\lambda^2$) which varies over a scale of $\xi_0$.  

It is also worth commenting on what kind of real systems could be described by our model.   Even if one neglects the weak Josephson coupling between the layers of a high-T$_c$ material, there will still be magnetic interactions between pancake vortices in adjacent layers, so that the system will not be truly 2D.  Likewise, if we wish to use this model to describe a very thin 2D high-T$_c$ film, there are stray fields in vacuum extending into the third dimension, which will contribute to the total energy.  In both cases, our model is definitely an oversimplification.   Nonetheless, our model does seem to describe some observed features in real high-T$_c$ materials, suggesting that it captures some significant physics in these systems.  Thus, we consider our model as a possible starting point for a fully realistic treatment of either 3D high-T$_c$ materials or very thin 2D 
high-T$_c$ layers.

We have connected our work on the pinning of vortices to continuing
efforts by several groups to describe the density of states near
vortices in cuprates superconductors.  To do this, we have
studied a model BCS Hamiltonian with d-wave symmetry, in which the
pairing field is obtained from simulations of the GL free energy
functional.  We use exact diagonalization to compute the local density of states 
on each atomic site of the
lattice described by this Hamiltonian.  For homogeneous systems, we found that
the LDOS near the vortex cores resembles that of a gapless
tight-binding model in two dimensions, with a Van Hove peak at zero
energy. However, when we introduce the inhomogeneities with a large
pairing field and large penetration depth that pin the vortices, the
LDOS near the vortex cores is markedly different from that of the
homogeneous systems.  Namely, this LDOS exhibits a large gap as well as
low and broadened peaks at energies greater than that of the
superconducting gap fa.r from the vortex core.  Also, our calculated
LDOS near the vortex cores in this inhomogeneous case does not exhibit
the spurious zero-energy peak which is present in several other
theoretical studies but is absent from experiments. All of these
features in our calculations are consistent with results observed in
STM experiments on YBCO \cite{maggio_renner_fischer,
  renner_revaz_fischer} and Bi2212 \cite{levy_kugler_fischer,
  pan_hudson_gupta, matsuba_sakata_kosugi, hoogenboom_renner_revaz}.

Our results are also consistent with those obtained of
Ref.\ \cite{melikyan_tesanovic}, using a somewhat
different model.  Those authors obtain better agreement between their 
calculated LDOS spectra and experiment\cite{pan_hudson_gupta}, if they assume
an enhanced rather than uniform pairing field near the vortex cores. In particular,
introducing this large pairing field near the vortex core in their model suppresses
the unphysical zero-energy peak.  Our approach, in which a large
gap is correlated with a large penetration depth, may 
heop justify the occurrence of this larger pairing field in the vortex cores.

Finally, we briefly comment on the possible physical origin of correlation between large
gap and large penetration depth.  The origin of the spatial fluctuations of $T_{c0}$ observed
in the cuprates may be the local fluctuations in concentration, which are highly likely since
the cuprates are mostly disordered alloys and the local concentration would involve an average
a coherence length, which is only about 15 $\AA$ .  But $T_{c0}$ is proportional to the gap (i.\ e.,
presumably, the pseudogap), which increases with decreasing concentration of charge carriers, whereas
$1/\lambda^2(0)$ is proportional to the superfluid density, which should decrease with decreasing
charge carrier concentration.  Therefore, we expect $T_{c0}$ to be positively correlated with
$\lambda^2(0)$, as seen experimentally and as used in the present model.

\section{Acknowledgments}  
This work was supported by NSF grant DMR04-13395.  The computations described here 
were carried out through a grant of computing time from the Ohio Supercomputer Center.
We thank Prof.\ J.\ Orenstein for a useful conversation about the
origin of the correlation between superconducting gap and penetration depth.




\newpage


\end{document}